# The Benefits of Hydrogen Energy Transmission and Conversion Systems to the Renewable Power Grids: Day-ahead Unit Commitment

Jin Lu
Department of Electrical and Computer Engineering
University of Houston
Houston, TX, USA
jlu28@uh.edu

Xingpeng Li
*Senior Member, IEEE*
Department of Electrical and Computer Engineering
University of Houston
Houston, TX, USA
xli82@uh.edu

*Abstract*—The curtailment of renewable energy is more frequently observed as the renewable penetration levels are rising rapidly in modern power systems. It is a waste of free and green renewable energy and implies current power grids are unable to accommodate more renewable sources. One major reason is that higher power transmission capacity is required for higher renewable penetration level. Another major reason is the volatility of the renewable generation. The hydrogen mix or pure hydrogen pipeline can both transfer and store the energy in the form of hydrogen. However, its potential of accelerating renewable integration has not been investigated. In this paper, hydrogen pipeline networks, combined with power-to-hydrogen (P2H) and hydrogen-to-power (H2P) facilities, are organized to form a Hydrogen Energy Transmission and Conversion System (HETCS). We investigate the operation of power systems coupled with HETCS, and propose the day-ahead security-constrained unit commitment (SCUC) with HETCS. The SCUC simulation is conducted on a modified IEEE 24-bus power system with HETCS. Simulation results show HETCS can substantially reduce the renewable curtailment, $CO_2$ emission, load payment and total operational cost. This study validates the HETCS can be a promising solution to achieve net-zero renewable grids.

## NOMENCLATURE

*Indices*
| | |
|---|---|
| $e$ | Electrolyzer. |
| $f$ | Fuel cell. |
| $g$ | Generator. |
| $k$ | Branch. |
| $n$ | Bus. |
| $t$ | Time period. |
| $w$ | Wind power plant. |

*Sets*
| | |
|---|---|
| $E$ | Set of electrolyzers in the power system. |
| $E(n)$ | Set of electrolyzers on bus $n$. |
| $F$ | Set of fuel cells in the power system. |
| $F(n)$ | Set of fuel cells on bus $n$. |
| $G$ | Set of generators in the power system. |
| $G(n)$ | Set of generators on bus $n$. |
| $K$ | Set of branches in the power system. |
| $K(n+)$ | Set of branches that the starting bus is $n$. |
| $K(n-)$ | Set of branches that the ending bus is $n$. |
| $T$ | Set of time intervals in a day. |
| $W$ | Set of wind power plants in the power system. |
| $W(n)$ | Set of wind power plants located at bus $n$. |

*Parameters*
| | |
|---|---|
| $c_g$ | Operational cost for generator $g$. |
| $c_g^{NL}$ | No load cost for generator $g$. |
| $c_g^{SU}$ | Startup cost for generator $g$. |
| $d_{nt}$ | Demand on bus $n$ at period $t$. |
| $P_e^{max}$ | Maximum capacity of electrolyzer $e$. |
| $P_f^{max}$ | Maximum capacity of fuel cell $f$. |
| $P_g^{max}$ | Maximum capacity of generator $g$. |
| $P_g^{min}$ | Minimum output power of generator $g$. |
| $P_k^{max}$ | Maximum thermal capacity of branch $k$. |
| $P_{wt}$ | Available wind active power at period $t$. |
| $R_g^{10}$ | Outage ramping limit in 10 minutes for generator $g$. |
| $R_g$ | Ramping limit in an hour for generator $g$. |
| $x_k$ | Reactance of branch $k$. |
| $\eta_e$ | Energy conversion efficiency of electrolyzer $e$. |
| $\eta_f$ | Energy conversion efficiency of fuel cell $f$. |

*Variables*
| | |
|---|---|
| $E_t$ | The total hydrogen stored in the hydrogen system. |
| $E_{nt}$ | The hydrogen energy stored at bus $n$. |
| $P_{et}$ | Active power consumed by electrolyzer $e$ at period $t$. |
| $P_{ft}$ | Active power generated by fuel cell $f$ at period $t$. |
| $P_{gt}$ | The output power of generator $g$ at period $t$. |
| $P_{kt}$ | Active power flow on branch $k$ at period $t$. |
| $P_{wt}^{Cur}$ | Active power curtailment for wind power plant $w$ at period $t$. |
| $r_{gt}$ | Reserve from generator $g$ in period $t$. |
| $u_{gt}$ | Commitment status for generator $g$ in period $t$. |
| $v_{gt}$ | Generator startup indicator, 1 if generator $g$ is turned on in period $t$; 0 otherwise. |
| $\theta_{kt}$ | Phase angle of the branch $k$ at period $t$. |

## I. INTRODUCTION

The renewable energy plays an increasingly important role in the modern power systems due to its sustainability and environmental friendliness. According to U.S. Energy Information Administration, renewable energy is expected to provide 24% of U.S. generation in 2023 [1]. While the increasing renewable energy leads to cleaner power grids, some challenges and issues arise due to intermittency of renewable generation. The flexible resources with high ramping rate and frequent start-up capability are required to ensure the reliability of the grids [2]-[3]. Besides, the increasing renewables cannot be well utilized if the grids do not suit for high renewable penetration [4]-[6]. Specifically, the transmission capacity of the power grids is limited, and unnecessary curtailment of wind/solar power is frequently observed when the network is congested [7]-[9]. For instance, California power grid was forced to curtail 187 GWh of wind and solar generation in 2015, and that number rose to 1,587 GWh in 2020 [10]-[11]. The curtailment of the renewables is a waste of the available green energy, and it also implies the barrier of deploying more renewables under the current grid condition. Furthermore, many potential benefits of the renewables such as accelerating the power system restoration [12] cannot be well utilized.

Lots of research are conducted on renewable grid integration by utilizing energy storage such as electric vehicles, batteries, compressed air and pumped-hydro storage [13]-[16]. The Power-to-hydrogen (P2H) technology can convert the electrical energy into hydrogen [17]-[18]. Since the hydrogen can be stored in the hydrogen storage facility and release the energy back to grid using hydrogen-to-power (H2P) facility [19]-[20]. The combination of P2H, hydrogen storage and H2P is capable to realize the energy storage function in the power grids. In [21], the performance of the energy hub (EH) that is equipped with an electrolyzer, hydrogen storage and a hydrogen turbine in the power system with high wind penetration level is investigated. It shows such combination is capable to reduce wind curtailments, especially for high wind energy penetration conditions.

A number of utilities around the world are upgrading or have set up plans to upgrade the natural gas pipeline network to a network that can handle larger portion of hydrogen mix or even pure hydrogen, while many other utilities are still conducting cost-benefit analysis for such upgrade [22]-[24]. Compared with local hydrogen storage, a hydrogen pipeline network can transmit the hydrogen in addition to storing the hydrogen. A Hydrogen Energy Transmission and Conversion System (HETCS) consisting of H2P, hydrogen pipeline network and P2H can interact and coordinate with the power system. The electrical energy can be converted into the hydrogen energy, stored in the hydrogen pipeline network and salt caverns, and transferred to other locations through the hydrogen pipeline network. Hence, a HETCS is expected to effectively address the over renewable generation and congestion-induced curtailment. To the best of the authors' knowledge, no current literatures investigate the feasibility and performance of a power system that is coupled with HETCS. The purpose of this paper is to fill this gap by studying the potential benefits of HETCS to the renewable power grids (RPG) with high penetration level of renewable generation.

Security-constrained unit commitment (SCUC) is to determine generator on/off status in power system day-ahead scheduling [25]-[27]. Existing SCUC only models the power system. In this paper, we propose a novel hydrogen-grid-based SCUC (H-SCUC) model for future HETCS and RPG coupled systems. Simulation is conducted on the modified IEEE 24-bus system coupled with a synthetic HETCS. The performance of HETCS-coupled power systems is analyzed and compared with (i) traditional power systems and (ii) EH-coupled power systems with local P2H and H2P but without hydrogen transmission.

The rest of this work is structured as follows. Section II explains a general day-ahead SCUC model. Section III presents the details of the HETCS and various SCUC formulations for different power systems with different configurations. Section IV presents the SCUC simulation results and analysis. The conclusions are drawn in Section V.

## II. GENERAL DAY-AHEAD SECURITY-CONSTRAINED UNIT COMMITMENT MODEL

The SCUC model is established to determine the optimal operational scheduling of the power system for the next day and is widely used in the industry. The optimal solution to SCUC is the most cost beneficial operational schedule, which includes generator commitment status and generation. The power system conditions such as transmission line power flow can also be obtained, as well as locational marginal price (LMP). Below is the formulation of a general SCUC model.

$$min \sum_{g \in G} \sum_{t \in T} (c_g P_{gt} + c_g^{NL} u_{gt} + c_g^{SU} v_{gt}) \quad (1)$$

$$P_g^{min} u_{gt} \leq P_{gt} \quad \forall g,t \quad (2)$$

$$P_{gt} + r_{gt} \leq P_g^{max} u_{gt} \quad \forall g,t \quad (3)$$

$$0 \leq r_{gt} \leq R_g^{10} u_{gt} \quad \forall g,t \quad (4)$$

$$\sum_{m \in G} r_{mt} \geq P_{gt} + r_{gt} \quad \forall g,t \quad (5)$$

$$P_{kt} = \theta_{kt}/x_k \quad \forall k,t \quad (6)$$

$$-P_k^{max} \leq P_{kt} \leq P_k^{max} \quad \forall k,t \quad (7)$$

$$\sum_{g \in G(n)} P_{gt} + \sum_{k \in K(n-)} P_{kt} - \sum_{k \in K(n+)} P_{kt} + \sum_{w \in W(n)} P_{wt}$$
$$- \sum_{w \in W(n)} P_{wt}^{Cur} = d_{nt} \quad \forall n,t \quad (8)$$

$$0 \leq P_{wt}^{Cur} \leq P_{wt} \quad \forall w,t \quad (9)$$

$$-R_g \leq P_{gt} - P_{g,t-1} \leq R_g \quad \forall g,t \quad (10)$$

$$v_{gt} \geq u_{gt} - u_{g,t-1} \quad \forall g,t \quad (11)$$

$$v_{gt} \in \{0,1\} \quad \forall g,t \quad (12)$$

$$u_{gt} \in \{0,1\} \quad \forall g,t \quad (13)$$

The object of the power grid daily operation model SCUC is to minimize the total cost of the power system, which consists of generator operation cost, generator no-load cost and generator startup cost. This objective function is represented in (1). Generators have maximum/minimum power limits, which is described by (2)-(3). The generator reserve requirements are considered using (4)-(5). The DC power flow equation is in (6) and the power flow limits is in (7). The nodal power balance equation is shown in (8). Since this study focuses on the curtailment of renewable energy, $P_{wt}^{Cur}$ is used to evaluate the curtailment of the wind power. The renewable curtailment limitation is in (9). The generator ramping constraint is shown in (10). The binary variables denoting generator status are $u_{gt}$ and $v_{gt}$, which are described by (11)-(13).

### III. DAY-AHEAD SECURITY-CONSTRAINED UNIT COMMITMENT FOR HETCS-COUPLED POWER SYSTEMS

#### A. Hydrogen Energy Transmission and Conversion System

This work is an initial investigation of feasibility and performance of utilizing hydrogen transmission network in power system operation. In this paper, we assume the hydrogen network consists of hydrogen pipelines between different nodes that are also interfaced with the power grid. The P2H and/or H2P facilities are placed at such locations covered by both the hydrogen network and the power transmission network. Hence, the hydrogen energy and electrical energy can be exchanged at these locations. The hydrogen network, P2H and H2P facilities constitute the HETCS that benefits the power system mainly through two functions: energy storage function (temporal flexibility) and energy transfer function (spatial flexibility).

i) Energy storage function: The electrical energy can be converted into the hydrogen energy by the P2H facilities. Since the hydrogen can be stored in the hydrogen network and other storage facilities such as salt caverns, the electrical energy can be stored in the form of hydrogen energy. The stored hydrogen energy can be converted back to the electrical energy by the H2P facilities.

ii) Energy transfer function: If a power system is coupled with HETCS, it can transfer energy through both the power transmission network and the hydrogen network to serve the loads. Specifically, the electrical energy of excess renewable generation can be converted into the hydrogen energy, transferred through the hydrogen network, and converted back to the electrical energy at the destination. Note that the energy can only be transmitted from where the P2H facilities are located to the H2P locations in the HETCS.

In HETCS-coupled power systems, the generated renewable energy can be converted into hydrogen energy and stored in the HETCS. Hence, the disadvantages of the intermittent renewable generation can be mitigated. Traditionally, a large amount of renewable generation requires higher transmission capacity. If the transmission lines are congested, it may cause the renewable curtailment. In HETCS-coupled power systems, the electrical energy generated by renewable sources can be transferred through HETCS in the form of hydrogen, which reduces the renewable curtailment. Based on the above reasons, we expect HETCS can improve renewable integration in power systems.

#### B. SCUC Formulation Considering HETCS

In this paper, the P2H facilities in the HETCS are electrolyzers, and the H2P facilities are the fuel cells. We use variable $P_{et}$ to represents the electrical power consumed by the electrolyzer $e$ at time $t$. Similarly, we use $P_{ft}$ to represents the electrical power generated by the fuel cell $f$ at time $t$. The nodal power balance equation considering the power consumed/generated by the electrolyzer/fuel cell is shown in (14). We use $\eta_e$ and $\eta_f$ to describe the energy conversion efficiency of the electrolyzer and fuel cell respectively. The hydrogen energy stored in HETCS can be calculated by (15). For any time interval, the hydrogen energy stored in the hydrogen network should be larger than zero (16). The electrolyzer/fuel cell power capacity constraints are shown in (17)-(18). Since this paper is focused on providing preliminary benefit analysis of the proposed HETCS to the renewable power grid, for simplicity, we ignored the costs associated with hydrogen transmission and storage.

$$\sum_{g \in G(n)} P_{gt} + \sum_{k \in K(n-)} P_{kt} - \sum_{k \in K(n+)} P_{kt}$$
$$+ \sum_{w \in W(n)} P_{wt} - \sum_{w \in W(n)} P_{wt}^{Cur} + \sum_{f \in F(n)} P_{ft}$$
$$= d_{nt} + \sum_{e \in E(n)} P_{et} \quad \forall n,t \quad (14)$$

$$E_t = E_{t-1} + \sum_{e \in E} \eta_e P_{et} - \sum_{f \in F} P_{ft}/\eta_f \quad \forall t \quad (15)$$

$$0 \leq E_t \leq E^{max}, \quad \forall t \quad (16)$$
$$0 \leq P_{et} \leq P_e^{max} \quad \forall e,t \quad (17)$$
$$0 \leq P_{ft} \leq P_f^{max} \quad \forall f,t \quad (18)$$

#### C. SCUC Models for Comparison

To gauge the operational performance of HETCS-coupled power systems, three benchmark cases are established for comparison: (i) a relaxed SCUC (R-SCUC) model that assumes infinite electrical network capacity, in which no transmission congestion will occur and HETCS is not included; (ii) a traditional SCUC (T-SCUC) that is a normal SCUC model for traditional power systems without HETCS; (iii) EH-coupled SCUC (EH-SCUC) model including energy hubs that have local electrolyzers, fuel cells and hydrogen storage to store/release the energy at the same location. The hydrogen energy stored in the tank can be calculated by (19). Moreover, the hydrogen energy stored in the hydrogen tank should always be positive (20). The formulations of various SCUC models are concluded in Table I.

$$E_{nt} = E_{n,t-1} + \sum_{e \in E(n)} \eta_e P_{et} - \sum_{f \in F(n)} P_{ft}/\eta_f \quad \forall n \in N^H, t \quad (19)$$
$$0 \leq E_{nt} \leq E_{nt}^{max}, \quad \forall n \in N^H, t \quad (20)$$

TABLE I
FORMULATIONS OF VARIOUS SCUC MODELS

| Proposed H-SCUC Model | R-SCUC Model | T-SCUC Model | EH-SCUC Model |
| --- | --- | --- | --- |
| Eq. (1)-(7) (9)-(18) | Eq. (1)-(6) (8)-(13) | Eq. (1)-(13) | Eq. (1)-(7) (9)-(14) (17)-(20) |

### IV. CASE STUDIES

The proposed H-SCUC model for HETCS-coupled power systems is tested, as well as the benchmark SCUC models, using the modified IEEE 24-bus system shown in Figure 1 [28]. In this test system, the HETCS consists of two electrolyzers and two fuel cells. The wind power plants are located on Bus 14 and

Bus 22. The electrolyzers are also located on these two buses. The fuel cells are located on Bus 13 and Bus 15, where peak loads are over 200 MW. The electrolyzer conversion efficiency $\eta_e$ is set to 0.8, and the fuel cell conversion efficiency $\eta_f$ is set to 0.6. For EH-coupled test case, two energy hubs are located on Bus 14 and Bus 22 where wind farms sit. Although we did not set a maximum limit on the hydrogen storage capacity for simplicity, we set maximum limits on the power capacities of electrolyzers and fuel cells. For a fair comparison, we set the same capacities of electrolyzers and fuel cells for both the HETCS-coupled power system and EH-coupled power system.

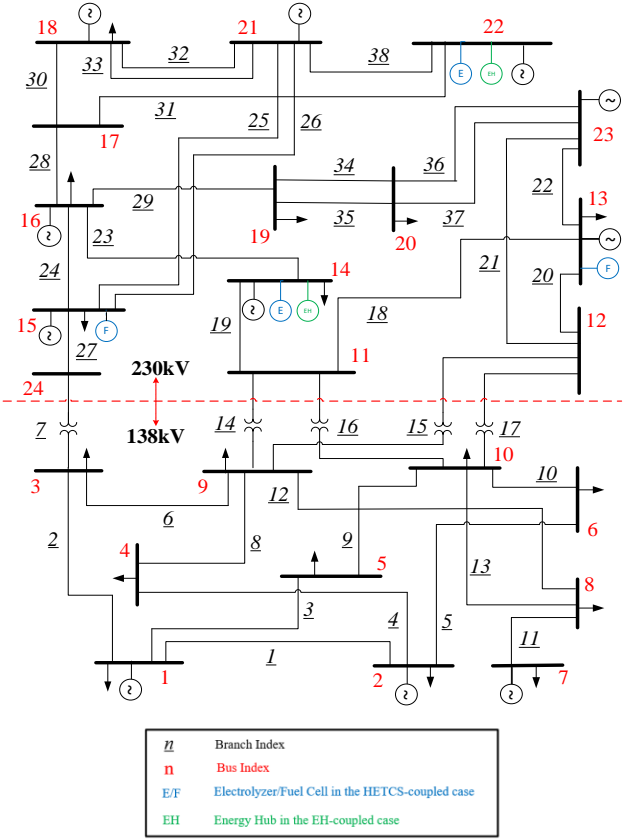

Fig. 1. The IEEE 24-bus system architecture (revised from [28]).

TABLE II
OPERATIONAL PERFORMANCE COMPARISON FOR 10% WIND PENETRATION

|  | R-SCUC Model | T-SCUC Model | EH-SCUC Model | H-SCUC Model |
| --- | --- | --- | --- | --- |
| Total Cost ($) | 1,126,615.1 | 1,135,979.3 (100%) | 1,128,001.2 (99.2%) | 1,126,805.1 (99.1%) |
| Total Load Payment ($) | 2,875,415.6 | 2,742,558.2 (100%) | 2,899,254.7 (105.7%) | 2,853,568.1 (104.0%) |
| Congestion Cost ($) | 0 | 9,364.1 (100%) | 1,386.0 (14.8%) | 190.0 (2.0%) |
| RC (MWh) | 0 | 0 (100%) | 0 (100%) | 0 (100%) |
| ANCLpH | N/A | 1.3 (100%) | 0.7 (53.8%) | 0.7 (53.8%) |
| NCLPH | N/A | 3 (100%) | 1 (33.3%) | 1 (33.3%) |
| CO2 Emission (Lbs×$10^6$) | 72.6 | 74.3 (100%) | 72.6 (97.7%) | 72.3 (97.3%) |

"RC" denotes Renewable Curtailment; "ANCLpH" denotes Average Number of Congested Lines per Hour; "NCLPH" denotes Number of Congested Lines in the Peak Hour.

The simulation is conducted on all four SCUC models at different wind penetration levels. We use python language with GLPK solving package to build and solve the cases. The simulation is performed on a desktop computer with Intel-i7 3.2 GHz CPU and 16 GB RAM. The SCUC simulation results for wind penetration level 10%, 30% and 50% are concluded in Table II, Table III and Table IV respectively. It is observed that HETCS can achieve cost saving, mitigate network congestion, lower load payment, and reduce the renewable curtailment and CO2 emissions substantially.

TABLE III
OPERATIONAL PERFORMANCE COMPARISON FOR 30% WIND PENETRATION

|  | R-SCUC Model | T-SCUC Model | EH-SCUC Model | H-SCUC Model |
| --- | --- | --- | --- | --- |
| Total Cost ($) | 706,582.3 | 922,125.5 (100%) | 712,789.2 (77.2%) | 700,928.3 (76.0%) |
| Total Load Payment ($) | 2,096,256.2 | 2,497,307.6 (100%) | 2,144,122.2 (85.8%) | 2,285,817.7 (91.5%) |
| Congestion Cost ($) | 0 | 215,543.2 (100%) | 6,206.8 (2.8%) | -5654.0 (-2.6%) |
| RC (MWh) | 0 | 5,343.0 (100%) | 335.8 (6.2%) | 0 (0%) |
| ANCLpH | 0 | 3.1 | 1.7 (54.8%) | 1.5 (48.3%) |
| NCLPH | 0 | 4 (100%) | 4 (100%) | 4 (100%) |
| CO2 Emission (Lbs×$10^6$) | 55.8 | 69.5 (100%) | 60.8 (87.4%) | 58.4 (84.0%) |

"RC" denotes Renewable Curtailment; "ANCLpH" denotes Average Number of Congested Lines per Hour; "NCLPH" denotes Number of Congested Lines in the Peak Hour.

TABLE IV
OPERATIONAL PERFORMANCE COMPARISON FOR 50% WIND PENETRATION

|  | R-SCUC Model | T-SCUC Model | EH-SCUC Model | H-SCUC Model |
| --- | --- | --- | --- | --- |
| Total Cost ($) | 437,160.3 | 842,091.2 (100%) | 632,472.1 (75.1%) | 561,891.6 (66.7%) |
| Total Load Payment ($) | 1,552,202.4 | 2,282,258.4 (100%) | 1,844,646.7 (80.8%) | 1,586,690.3 (69.5%) |
| Congestion Cost ($) | 0 | 404,930.9 (100%) | 195,311.8 (48.2%) | 124,731.3 (30.8%) |
| RC (MWh) | 0 | 14,096.6 (100%) | 7,291.4 (51.7%) | 5,007.3 (35.5%) |
| ANCLpH | 0 | 3 (100%) | 3 (100%) | 3 (100%) |
| NCLPH | 0 | 3 (100%) | 4 (133.3%) | 4 (133.3%) |
| CO2 Emission (Lbs×$10^6$) | 36.2 | 68.4 (100%) | 58.8 (85.9%) | 52.8 (77.1%) |

"RC" denotes Renewable Curtailment; "ANCLpH" denotes Average Number of Congested Lines per Hour; "NCLPH" denotes Number of Congested Lines in the Peak Hour.

The wind power curtailment of different cases is plotted in Figure 2. From the plots, we can observe that the wind curtailment increases when the wind penetration level is high. The wind curtailment of the HETCS-coupled case is much lower than the traditional case and the EH-coupled case.

The carbon emission is calculated based on the scheduled generator output power in the SCUC results. The emission data of the generators is from [29]. The carbon emission for different wind penetration levels is plotted and shown in Figure 3. We can observe that HETCS-coupled case can significantly reduce the carbon emission. It reduces more carbon emission than EH-coupled case when the wind penetration level is high.

The total operational cost of different cases is shown in Figure 4. When the wind penetration level is higher, the total operational cost is lower. From the plot, we can observe that the HETCS case has lower operational cost than the EH-coupled case and the traditional case. The HETCS case can obviously save more cost when the wind penetration level is higher. The total operation cost of the HETCS-coupled case is even lower than the non-congestion case when the wind penetration level is 30%. The reason is that the HETCS can provide the function of the energy storage, while no energy storage is included in the non-congestion case.

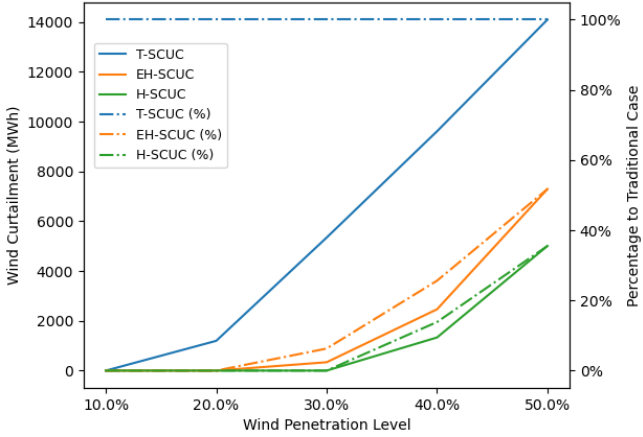
Fig. 2. Wind Curtailment for Different Wind Penetration Level.

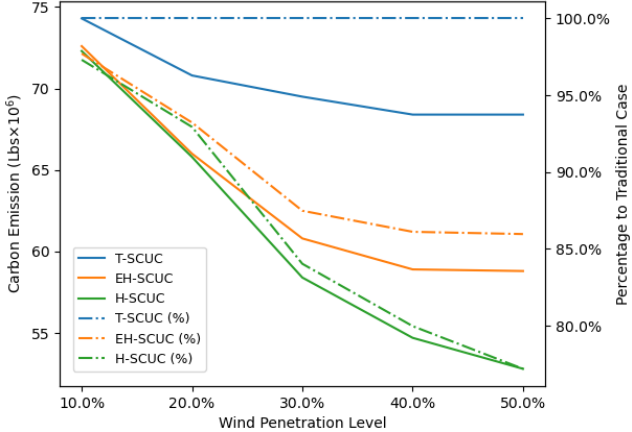
Fig. 3. Carbon Emission for Different Wind Penetration Level.

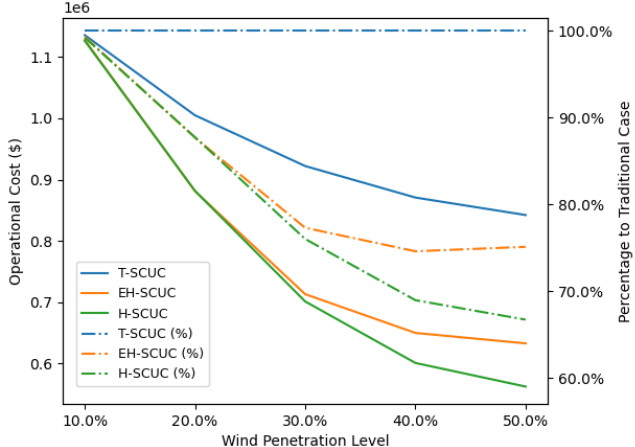
Fig. 4. Total Operational Cost for Different Wind Penetration Level.

The average LMPs of different cases is shown in Figure 5. The average LMP of the HETCS-coupled case is lower than the EH-coupled case for most wind penetration scenarios. Besides, the average LMP of HETCS-coupled case is much lower than the traditional case.

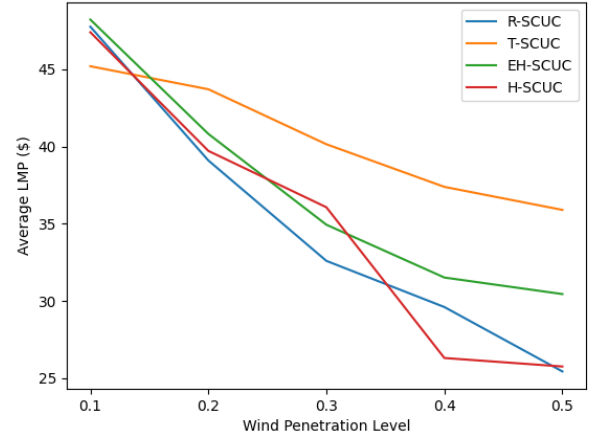
Fig. 5. Average LMP for Different Wind Penetration Level.

TABLE V
WIND CURTAILMENT OF DIFFERENT FUEL CELL SITE SELECTION CASES UNDER DIFFERENT WIND PENETRATION LEVELS

|  | 10% Wind (MWh) | 20% Wind (MWh) | 30% Wind (MWh) | 40% Wind (MWh) | 50% Wind (MWh) |
|---|---|---|---|---|---|
| EH-coupled Case (Benchmark) | 0 | 0 | 335.8 (100%) | 2544.5 (100%) | 7291.4 (100%) |
| Site Selection Case 1 | 0.0 | 0.0 | 0.0 | 1331.8 (52.3%) | 5007.3 (68.6%) |
| Site Selection Case 2 | 0.0 | 0.0 | 0.0 | 1332.7 (52.3%) | 4270.8 (58.5%) |
| Site Selection Case 3 | 0.0 | 0.0 | 0.0 | 1339.1 (52.6%) | 4303.2 (59.0%) |

TABLE VI
TOTAL COST OF DIFFERENT FUEL CELL SITE SELECTION CASES

|  | 10% Wind | 20% Wind | 30% Wind | 40% Wind | 50% Wind |
|---|---|---|---|---|---|
| EH-coupled Case (Benchmark) | $1.128M (100%) | $0.882M (100%) | $0.712M (100%) | $0.651M (100%) | $0.632M (100%) |
| Site Selection Case 1 | $1.126M (99.8%) | $0.881M (99.9%) | $0.700M (98.3%) | $0.600M (92.0%) | $0.561M (88.8%) |
| Site Selection Case 2 | $1.126M (99.8%) | $0.880M (99.8%) | $0.699M (98.1%) | $0.597M (91.6%) | $0.549M (86.8%) |
| Site Selection Case 3 | $1.126M (99.8%) | $0.881M (99.8%) | $0.700M (98.2%) | $0.594M (91.1%) | $0.544M (86.1%) |

The P2H and H2P facilities should be placed at proper locations to ensure the HECTS-coupled power system can perform well. To find the suitable locations of fuel cells, we conduct the H-SCUC simulations on the HETCS-coupled systems with different fuel cell locations. Based on the load distribution of the IEEE 24-bus system, three fuel cell site selection cases are compared: (i) the fuel cells are located at the high-loaded buses. Bus 13 and Bus 15 are selected where peak loads are over 200 MWh. It's the HETCS-coupled case in the various SCUC comparison; (ii) the fuel cells are located at the center of load area. In the IEEE-24 bus system, the 138kV area has more loads than the 230kV area. Bus 4 and Bus 5 are selected in this case; (iii) the fuel cells are located at the bridges between the generation area and the load area. Bus 9 and Bus 10 are selected in this case. For fair comparison, the electrolyzers are located at Bus 14 and Bus 22 where wind farms sit in all above cases. Besides, we select the EH-coupled case which the fuel cells and electrolyzers are located at Bus 14 and Bus 22 as the benchmark case. The simulation results are shown in Table V and Table VI.

The simulation result shows that different fuel cell locations may not influence the HETCS-coupled power system performance when wind penetration is low. When wind penetration is 40% or more, the proper selection of fuel cells can considerably reduce renewable curtailment and the total cost. Compared to random selection of high-loaded buses, locating the fuel cells at the center of load area or the bridges between the load area and generation area is more likely to attain better performance of HETCS-coupled systems. Determining the best P2H and H2P locations is a complicated problem, and it needs further thorough investigation in future.

## V. Conclusion

Power systems with high renewable penetration can couple with the hydrogen energy conversion and transmission system to gain better operational performance. The HETCS consists of hydrogen network, P2H, and H2P facilities. It can both transfer and store the energy in the form of hydrogen. The unit commitment model of the HETCS-coupled power system is proposed to study its operational performance. To better analyze the benefits of the HETCS-coupled system, the non-congestion case, traditional case and EH-coupled case are also developed for comparison. The simulation result shows that the HETCS can substantially reduce the renewable curtailment once the renewable penetration is 20% or above. Other operational performance such as the total operational cost, LMPs and carbon emission also shows that HETCS is a promising solution for power systems with high renewable penetration levels. Since this study focuses on the feasibility and benefits of the HETCS-coupled power system, the modeling of the HETCS is brief. Detailed HETCS modeling in SCUC and other applications in practical power systems is required in future work.